\documentclass{ws-procs9x6}

\begin{document}

\title{Status of KASKA: \\
The Japanese Reactor $\sin^22\theta_{13}$ Project\\}

\author{F.Suekane (\lowercase{for the} KASKA
C\lowercase{ollaboration}\footnote{\uppercase{T}he \uppercase{KASKA}
group members at the 
time of the conference were,  
\uppercase{N.T}amura, \uppercase{M.T}animoto, \uppercase{H.M}iyata,
\uppercase{R.W}atanabe (\uppercase{N}iigata \uppercase{U}niv.), 
\uppercase{Y.S}akamoto, (\uppercase{R}ikkyo \uppercase{U}niv.),
\uppercase{F.S}uekane, \uppercase{K.I}noue, \uppercase{T.A}raki, 
(\uppercase{T}ohoku \uppercase{U}niv.),
\uppercase{M.K}uze, (\uppercase{T}okyo \uppercase{I}nsititute of 
\uppercase{T}echnology),
\uppercase{T.S}umiyoshi, \uppercase{H.M}inakata, \uppercase{O.Y}asuda and 
\uppercase{H.S}ugiyama (\uppercase{T}okyo \uppercase{M}etropolitan 
\uppercase{U}niv.)} ) \\} 

\address{suekane@awa.tohoku.ac.jp\\
Research Center for Neutrino Science, \\
Graduate School of Science, Tohoku University, \\
Sendai, 980-8578, Japan}  


\maketitle  

\abstracts{ Last year, in the NOON03 conference, we pointed out the importance of reactor 
$\sin^22\theta_{13}$ measurement and showed a realistic experimental idea to measure 
$\sin^22\theta_{13}$ precisely using world's most powerful reactor complex.  
Since then, the conceptual design of the experiment and negotiations
with electric power company, have been progressing. 
In this proceedings for the NOON04 conference, the present status of the
proposed experiment, now called KASKA, is described.  
  }

\section{Introduction}
Since neutrinos have peculiar properties such as very small masses,
very much different mixing patterns from quark sector, 
possibility of being Majorana particle, study of neutrinos is very
important to understand the nature of the elementary particles and to construct its unified
theory in the future.
Because the neutrinos have finite masses and mixing, the neutrinos perform
neutrino oscillation if their masses are not equal. 
For two flavor case, the probability that a 
particular flavor neutrino $\nu_x$ 
with energy $E_\nu$ to remain as $\nu_x$ after traveling the distance
$\it{L}$ becomes,
\begin{equation}
P_{\nu_x \rightarrow \nu_x} =1-\sin^22\theta \sin^2\frac{\Delta m^2 L}{4E_{\nu}},
\end{equation}  
due to the neutrino oscillation, where, $\Delta m^2$ is the difference of
the squared mass of the two 
mass eigen states; $ m^2_2-m^2_1$, and
$\theta$ is the mixing angle.
Through neutrino oscillation, it is possible to access to a very small
mass range where direct measurements are difficult to reach.   
Experimental studies of the neutrino oscillations have been
progressing very rapidly in these days.  
The first firm evidence of the neutrino oscillation was discovered in the
disappearance of muon type atmospheric neutrinos by
SuperKamiokande group in 1998\cite{SK}. 
Recently the K2K group is confirming the result\cite{K2K} by observing the
disappearance of $\nu_{\mu}$ produced by the KEK-PS accelerator.
These phenomena are naturally explained by the $\nu_{\mu} \rightarrow
\nu_{\tau}$ oscillation and the measured oscillation parameters are
$\Delta m^2_{23} \approx 2 \times 10^{-3} eV^2$ and $\sin^22\theta_{23}
\approx 1$. 
As for the electron type neutrinos there have been indications of the
neutrino oscillation in the deficit of the solar neutrinos \cite{Solar}
for long time. 
A transformation of the solar $\nu_e$ to other type neutrinos was
identified by SNO group in 2002 \cite{SNO}. 
The KamLAND group observed a large deficit in reactor $\bar{\nu}_e$ in
2002\cite{KamLAND}. 
Combined analysis shows that the oscillation parameters are
$\Delta m^2_{12} \approx 7 \times 10^{-5} eV^2$ and $\sin^22\theta_{12}
\approx 0.8$.  
The remaining important subjects of the neutrino physics is to 
measure finite value of or to set a strong limit on the last mixing angle
$\theta_{13}$, and then 
to measure the leptonic CP violating phase $\delta_{CP}$.
The $\theta_{13}$ can be measured by either reactor $\bar{\nu}_e$
disappearance at the baseline around 1.5km or
$\nu_{\mu} \rightarrow \nu_e$ appearance experiment at around $\Delta
m^2 \approx 2 \times 10^{-3}eV^2$. 
   The most stringent upper limit of $\theta_{13}$ was measured by CHOOZ
group using reactor 
$\bar{\nu}_e$s to be $\sin^22\theta_{13}<0.2$ at $\Delta m^2=2 \times
10^{-3}eV^2$~\cite{CHOOZ}. 
Long baseline accelerator experiments (LBL) which has sensitivity to
$\theta_{13}$ have been approved~\cite{MINOS}~\cite{JPARC}.
On the other hand, improving the CHOOZ limit significantly
by using reactor neutrinos with near/far detector strategy had been
considered by Krasnoyarsk group~\cite{Kr2Det}. \\ 
In late 2002, we pointed out in the paper~\cite{minakata} that the reactor
$\theta_{13}$  measurement is important despite some LBL $\theta_{13}$
experiments would run in the near future.   
The point of this paper was that if the Large Mixing Angle (LMA)  
solution for the solar neutrino problem is correct, 
the  $\sin\delta_{CP}$ terms in eq.-(\ref{eq:LBL}) can not be ignored any more 
(this very reason enables the measurement of $\delta_{CP}$ in future LBL
experiments)
and it is difficult to extract $\theta_{13}$ information from the
measurement of $\nu_e$ appearance.

\begin{equation}
 \begin{split}
P(\nu_{\mu} \rightarrow \nu_e)\approx&\sin^22\theta_{13}\sin^2\theta_{23}  \\
&-\frac{\pi}{2}\frac{\Delta m^2_{12}}{\Delta m^2_{23}}\cos\theta_{13}
\sin2\theta_{12}\sin2\theta_{23}\sin2\theta_{13}\sin\delta_{CP} 
 \end{split}
\label{eq:LBL}
\end{equation}

Furthermore the degeneracy problem of  $\theta_{23}$ disturbs the
$\theta_{13}$ determination by LBL experiments.   
On the other hand the reactor based $\bar{\nu}_e$  oscillation
measurement is a pure
$\sin^22\theta_{13}$  measurement and by combining with the LBL
measurement, there is a possibility to obtain precious
information such as solving such degeneracies and even obtaining a clue
of non-0 $\delta_{CP}$ before the 
$\bar{\nu}$ operation mode of LBL experiments.  
A realistic idea to measure $\sin^22\theta_{13}$ precisely using
$\bar{\nu}_e$ coming from the Kashiwazaki-Kariwa nuclear power station
was also shown in the paper.  
Meanwhile, KamLAND showed that the LMA is the correct answer and that 
both $\Delta m^2_{12}$ and $\sin^2 2\theta_{12}$ are reasonably large
that there is a possibility to measure $\sin\delta_{CP}$ in the future
experiments.  
Because detectability of $\sin\delta_{CP}$ depends also on $\theta_{13}$,
the reactor $\sin^22\theta_{13}$ measurement became important
accordingly and enthusiasm to such experiments broke out.  
In NOON03 (Feb. 2003), there were three talks on reactor-$\theta_{13}$
experiment from Japan~\cite{suekane}, USA~\cite{shaevitz} and
Europe~\cite{thierry}.  
There have been 3 workshops dedicated to the reactor-$\theta_{13}$
within a year, in Alabama, USA (April/'03), in Munich, Germany
(October/'03) and in Niigata, Japan (March/'03).
There has been a lot of activities in worldwide.  
Along with these activities, a white paper for the reactor-$\theta_{13}$
measurement was written by 125 authors~\cite{white} in January, 2004.   
The conclusion of the white paper stresses the importance of the
reactor-$\theta_{13}$ measurement and shows that it is possible to build
such detectors which have sensitivity $\sin^22\theta_{13}=0.02$ or
better by making use of current neutrino detection technologies
with a relatively short time period. 
The cost of this kind of experiment is small if compared with
LBL experiments and these measurements can be done by very cost-effective way.
Now there are 7 proposed sites all over the world; from Japan, Europe,
USA, Russia, China and Brazil. 
The status of the Japanese KASKA project is described below. 
\\  

\section{KASKA experiment}

\begin{figure}
\centerline{\epsfxsize=3.6in\epsfbox{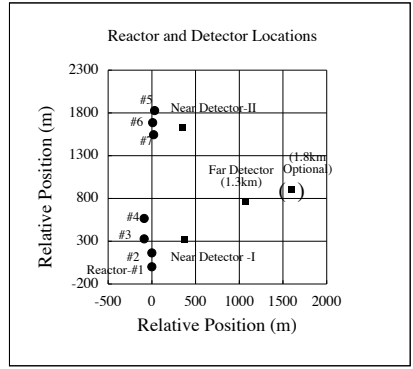}}
\caption{
The reactor and planned detector positions. Kashiwazaki-Kariwa
Nuclear power station consists of seven powerful reactors forming two clusters
each has 4 and 3 reactors respectively. 2 near detectors are placed at
 around 400m from corresponding cluster. The distance between the far detector
 and the reactors are around 1.3km. There is an optional plan to put
 the detector at 1.8km. 
}
\label{fig:site}
\end{figure}
\begin{figure}[ht]
\centerline{\epsfxsize=4.0in\epsfbox{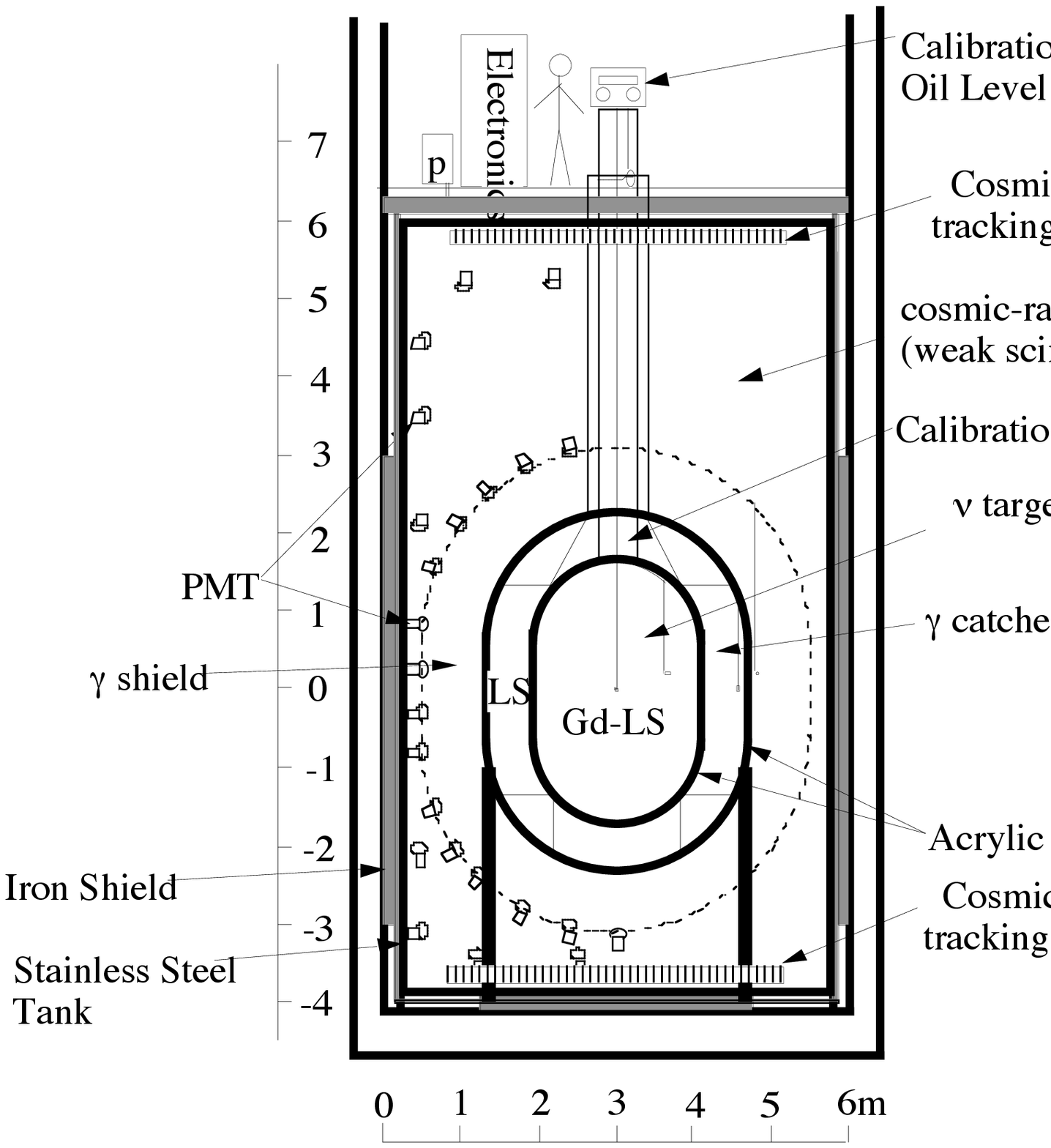}}
\caption{Schematic view of the detector. 
The $\bar{\nu_e}$ target is 8ton Gd loaded liquid scintillator. 
The Gd concentration is 0.1\% or more, corresponding to the neutron
 absorption probability on Gd of 88\% or more. 
The $\bar{\nu_e}$ target is surrounded by 60cm thick $\gamma$ catcher
 scintillator.  
This scintillator catches $\gamma$ rays which escape from the Gd loaded
 liquid scintillator and to  reconstructs original neutron absorption and
 positron energy.   
The $\gamma$ catcher scintillator is surrounded by 90cm thick buffer oil. 
This buffer oil shield $\gamma$-rays from PMT glass, from which CHOOZ
 experiments suffered much. 
The outer most layer is muon anti-counter made of weak scintillator to
 improve cosmic-ray tagging efficiency.  
The far detector will be placed at the bottom of 200m shaft hole with
 diameter 6m.  
The near detectors will be placed at the bottom of 70m depth shaft hole.   
\label{fig:detector}}
\end{figure}

KASKA is an abbreviation of $\it{Kas}$hiwazaki-$\it{Ka}$riwa Nuclear
Power Station, following the tradition to name experiment after the
power station to use.  
The Kashiwazaki-Kariwa nuclear power station, which is the world's most
powerful nuclear reactor complex, has seven reactors and generates 
thermal energy of 24.3GW. 
It produces $4.4 \times 10^{22} \bar{\nu}_e$/second when all the
reactors are up.   
The seven reactors are
arranged approximately in line forming two clusters, each consisting of 3 and 4
reactors as shown in fig.-\ref{fig:site}.    
In KASKA experiment, three identical neutrino detectors will be used.
Two near detectors are placed at an approximate distance of 400m from
corresponding cluster and the far detector is placed at the distance of
1.3km (1.8km optional) from all the reactors.  
By comparing the data of near-far detectors, most systematics,
such as detector efficiency and neutrino flux ambiguity will cancel out.  
The detectors are placed deep underground using shaft holes with
diameter 6m. 
The depth of the shaft hole is 200m (250m for 1.8km case) for far
detector and 70m for two near detectors.
The cosmic ray rate is estimated to be 0.35$/m^2/s$ for the far detector
and ten times more for the near detectors\cite{MUSIC}. 
The cosmic-ray directly falls from the open space above the
detector is only a fraction of the total cosmic-ray rate.    
Fig.-~\ref{fig:detector} shows the schematic view of the detector. 
The central part (Region-I) is 8tons of Gd loaded liquid scintillator (Gd-LS)
with Gd concentration of 0.1\% or more. 
This Gd-LS is the PaloVerde type, which is stable for 
at least a few years in an acrylic container~\cite{PaloVerdeLS}. 
The reactor $\bar{\nu}_e$ is detected with the following process. 
\begin{equation}
\bar{\nu}_e + p \rightarrow e^+ + n
\end{equation}
Then the neutron is absorbed by the Gd, producing $\gamma$-rays whose
total energy amounts to 8MeV.
\begin{equation}
  n+Gd \rightarrow Gd' +\gamma s (\sum E_{\gamma}=8MeV)
\end{equation}
By taking the coincidence of the positron and neutron signals,
backgrounds are severly suppresed. 
60,000events (30,000events with 1.8km baseline) will be collected in
the far detector with three years of operation, assuming the reactor and
detector efficiency to be 50\%. 
The corresponding statistic error is 0.4\% (0.6\%). 
The near detectors will collect 600,000 events in the same period.    
The Gd-LS is contained in an acrylic container with the shape of two
half spheres sandwiching a cylinder.  
The radius of the sphere and the cylinder is 1.1m. 
The height of the cylinder part is 1.2m and the volume is $10m^3$.
The $\bar{\nu}_e$ event rate is roughly 100/day for far detector and 5
times more for each near detector.   
There is a normal liquid scintillator layer (Region-II) outside of the
Gd-LS, having the same light output as the Gd-LS.  
This region is used to catch the $\gamma$-rays which escape from
Region-I and to reconstruct the original energies of neutron and positron
signals.  
The region-II scintillator is contained in the 2nd acrylic container. 
There are 90cm thick non-scintillating buffer oil (Region-III) between
Region-II and PMT.  
This layer is used to absorb the $\gamma$-rays form the PMT glass. 
There are 400 8 inch PMTs at the outside of the Region-III, whose
photo-cathode covers 10\% of total area.  
The glass of PMT will be a low background type. 
With the buffer oil, the single rate from the PMT glass will reduce to
be less than 5Hz with 0.9MeV energy threshold. 
The outer most region (Region-IV) is the cosmic-ray anti-counter. 
A high efficiency of cosmic-ray anti counter is essential to reduce the
cosmic-ray associated background and weak scintillator will be used. 
The upper part of the anti counter is made significantly thick to
efficiently reduce the fast neutron background.  
There are cosmic-ray tracking devices at the top and bottom of the
anti-counter region.  
The position resolution will be around 10cm, corresponding to the vertex
reconstruction resolution of the neutrino events.
The measured cosmic-ray track is used to estimate spallation background
such as $^9$Li based on distribution of the distances between
the event and the track.
The absolute contamination values of such correlated spallation events
are expected to be order of 0.5\% and can be measured  to the accuracy
of better than 0.3\%.
The spallation single events are used as the energy and position
calibration.  
All the liquid is contained in double layers of stainless steel tank and
steel tank for safety.  
The steel tank is also used as the geomagnetic shield.  
Outside of these tanks, there is thick iron layers to shield gamma rays
from the soils outside.  
The nature of the soil around the detector is a watery mud of density of
1.75g/cc. 
This low density and high water component are expected to help to reduce
the fast neutron background. 
At the top of the detector there will be additional thick iron layer to
absorb hadronic and soft components of the cosmic rays which come
directly from the open shaft area at the top.  
The readout electronics is being designed based on 500MHz-10bit flash
ADC and FPGA.  

Because the single background rate will be only a few Hz at 1MeV positron
threshold and less than 0.01Hz for 4MeV neutron threshold, the
accidental background is less than 1\% and this can be precisely
measured by shifting the coincidence window. 
The main background will be fast neutrons. 
The absolute amount of the neutron background is expected to be 1\% and
it can be measured with precision of 20\% using the prompt energy
spectrum at above the reactor $\bar{\nu}_e$ range. 

The event selections are simple. (1) $0.9MeV<E_{pronpt}$, (2)
$4MeV<E_{delayed}$  and (3) $ 1\mu s <t_{delay}-t_{prompt}<200\mu s$. 
Because the prompt signal energy is larger than 1.022MeV, and the delayed
energy peaks at 8MeV, the selection (1) and (2) are insensitive to the
energy calibration error. 
The neutron capture signal on $^{12}$C (E=4.95MeV, 0.05\% of neutron
absorption) can be used as an energy calibration point for the cut (2). 
The selection (3) makes use of relative timing and can be performed
precisely. 
Because there is no fiducial cut, it is free from the error associating
with position reconstruction. 

The main error of the neutrino detection efficiency comes from the
relative difference of the volume of the Gd-LS between the detectors.  
The volume will not be defined by the container volume because it
possibly deforms under the pressure of the liquids.  
But it will be defined by the volume of the LS which is measured precisely
when introducing in the detector.    
The expected accuracy of the measurement will be 0.5\% or better.

\begin{figure}[ht]
\centerline{\epsfxsize=3.0in\epsfbox{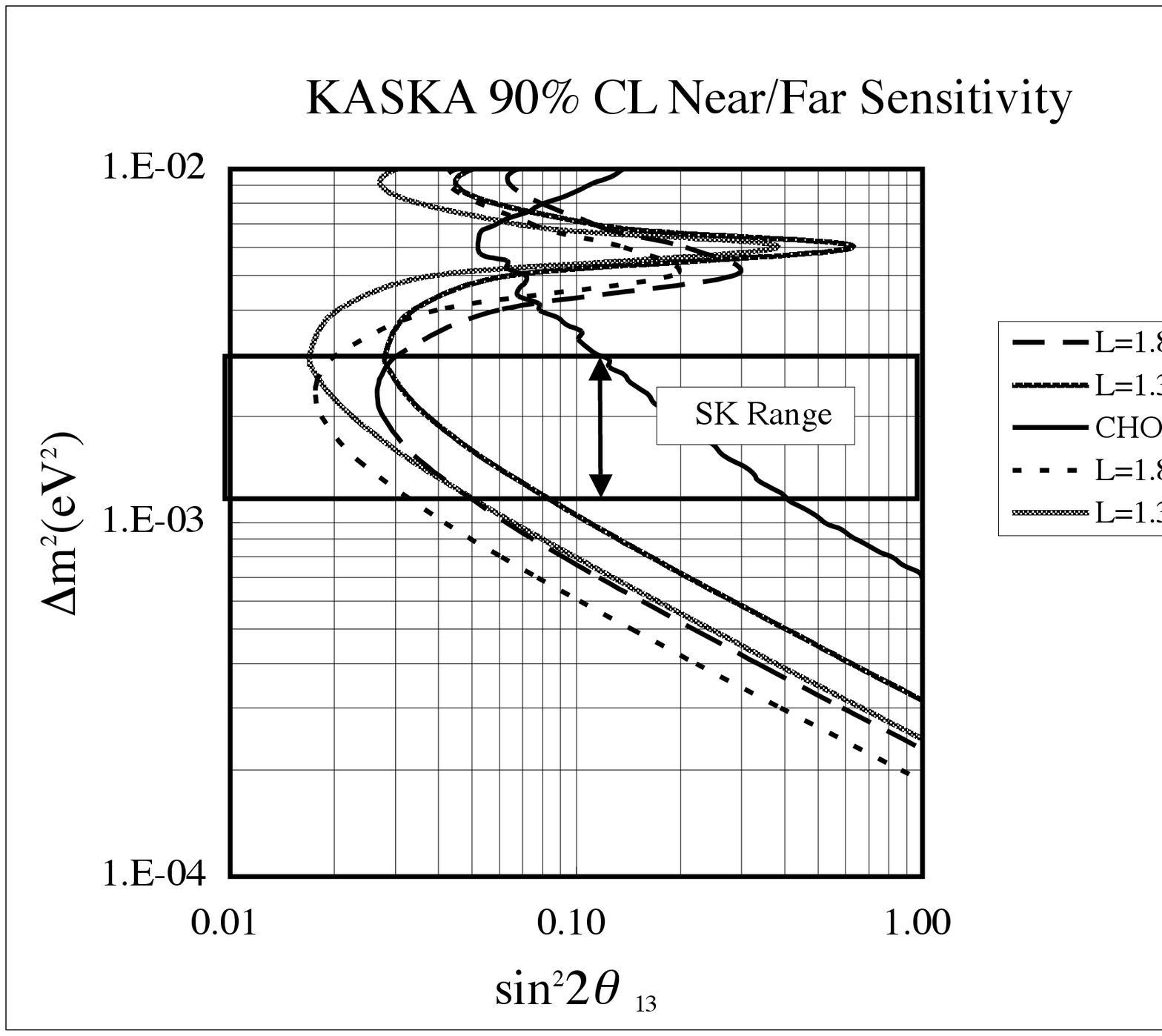}}
\caption{ KASKA 90 \% C.L. sensitivity for the case of far detector
 locations of 1.3km and 1.8km and the systematic uncertainty of 0.5\%
 and 1\%. The average distance of the near detectors is 400m.  
The recent SuperKamiokande results of atmospheric neutrino oscillation. 
The dip at $\Delta m^2 \approx 6 \times 10^{-3} eV^2$ is due to the fact
 that the disappearance rate at near and far detectors become same at
 this $\Delta m^2$.  
In the actual analysis, the limit of this region is governed by the
 absolute systematic error of the detection efficiency and there will
 not be such deep valley. 
\label{fig:sensitivity}}
\end{figure}

Since there are significant improvements of the detector from CHOOZ
experiment, whose detector associated error is 1.5\%, we anticipate the
absolute error of single KASKA detector can be 1\% or less.   
By comparing near and far detectors most of the remaining systematics
and ambiguities associated to the neutrino flux 
cancel and the final systematic error is expected to be much better than 
1\%.  
Fig.-\ref{fig:sensitivity} shows the expected near/far sensitivity after
3 years of operation for the 
cases of 1\% and 0.5\% of systematic errors and 1.3km and 1.8km
baselines.  
It is possible to improve the sensitivity down to $\sin^22\theta_{13}=
0.017 \sim 0.027$. 

\section{Summary}
Now it can be said that it is consensus among the neutrino physicists that the 
reactor-$\theta_{13}$ measurement is important. 
The KASKA experiment will measure the last neutrino mixing angle
$\sin^22\theta_{13}$ with 90\%C.L. accuracy of $0.017 \sim 0.027$
using reactor $\bar{\nu}_e$.   
The reactor measurement of $\sin^22\theta_{13}$ with this sensitivity is
important because if finite $\sin^22\theta_{13}$ is observed by this
experiment, it means that there is a good chance to observe $\sin
\delta_{CP}$ in future experiments.  
If $\sin^22\theta_{13}$ is too small to be detected by this
experiment, probably it will become necessary to perform another set of 
reactor-$\theta_{13}$ experiments with higher sensitivity before LBL
experiments to detect CP violation, to check if there is still
possibility to measure $\sin \delta_{CP}$ in such experiments.  
Also it means that the $\nu_3$ component of the $\nu_e$ is really
small unlike other neutrino mixing, and exploring the reason of this
smallness may play a key roll when constructing the unified theory of
the elementary particles.\\
The conceptual design of the KASKA detectors is almost complete and
negotiations with the electric power company and various administrative
organs are progressing well.  
Recently, detector R\&D funds are approved and we are hoping that the project
is approved in the near future. 

%
%\section*{Acknowledgments}  

\end{document}